\documentclass[doublecol]{epl2}
\usepackage{graphics}
\usepackage{epsfig}
\usepackage{amssymb,amsfonts,graphicx}
\usepackage{rotating}
\usepackage{subfigure}
\usepackage{dcolumn}
\usepackage{bm}

\title{Information flow between stock indices}

\author{Okyu Kwon\inst{1,2} \and Jae-Suk Yang\inst{1}\thanks{Corresponding author.}}
\shortauthor{O. Kwon and J.-S. Yang}

\institute{
  \inst{1} Department of Physics, Korea University, Seoul 136-701,
Korea\\
  \inst{2} National Creative Research Initiative Center for
Neuro-dynamics, Korea University, Seoul 136-701, Korea }
\pacs{89.65.Gh}{Economics; econophysics, financial markets,
business and management} \pacs{89.70.Cf}{Entropy and other
measures of information} \pacs{89.75.Fb}{Structures and
organization in complex systems}

\abstract{Using transfer entropy, we observed the strength and
direction of information flow between stock indices. We uncovered
that the biggest source of information flow is America. In
contrast, the Asia/Pacific region the biggest is receives the most
information. According to the minimum spanning tree, the GSPC is
located at the focal point of the information source for world
stock markets.}

\begin{document}

\maketitle

\section{Introduction} \label{intro}

Economic systems have recently become an active field of research
for physicists striving to transfer concepts and methodologies
from statistical physics such as phase transition, fractal theory,
spin models, complex networks, and information theory to the
analysis of economic problems
\cite{eguiluz,krawiecki,chowdhury,takaishi,kaizoji02,kaizoji04,palagyi,kaizoji01,wsjung1,wsjung2,jsyang3,arthur,mategna,bouchaud,mandel,kullmann,giada,jbpark,jwlee,jsyang2,kaizoji03,matal,yang,silva}.

Among the numerous methodologies put forward, time series analysis
has proven to be one of the most efficient methods and is widely
applied to the examination of characteristics of stock and foreign
exchange markets. In order to analyze financial time series, a
range of statistical measures have been introduced, including
probability distribution
\cite{jsyang2,kaizoji03,matal,yang,silva,stanley1,mccauley},
autocorrelation \cite{yang}, multi-fractal \cite{kkim}, complexity
\cite{jbpark,jwlee,jsyang2}, entropy density \cite{jwlee,jsyang2},
and transfer entropy \cite{jsyang3}.

Information is a keyword in analyzing financial market data or in
estimating the stock price of a given company. It is quantified
through a variety of methods such as cross-correlation,
autocorrelation, and complexity. However, while they may be
appropriate measures for the observation of the internal structure
of information flow, they fail to illuminate the directionality of
information flow. Schreiber \cite{schreiber} introduced transfer
entropy, which measures the dependency in time between two
variables and notes the directionality of information flow. This
concept of transfer entropy has already been applied to the
analysis of financial time series. Marschinski and Kantz
\cite{marschinski} calculated information flow between the Dow
Jones and DAX stock indexes to better observe interactions between
the two huge markets. Kwon and Yang \cite{jsyang3} measured the
direction of information flow between the composite stock index
and individual stock prices, while the information flow among
individual stocks in a stock market has been estimated by Baek and
coauthors \cite{baek} in order to measure the internal structure
of a stock market.

\begin{table}[tbp]
\centering%
\begin{tabular}{l|l|l|l}
\hline
Americas & 1&MERV & Argentina\\
&2 & BVSP& Brazil\\
&3 & GSPTSE& Canada \\
& 4& MXX&Mexico \\
& 5& GSPC&U.S.  \\
& 6& DJA&U.S.  \\
& 7& DJI&U.S.  \\
\hline
Asia/Pacific& 8& AORD&Australia  \\
& 9&SSEC &China \\
& 10& HSI& China\\
&11& BSESN&India  \\
& 12& JKSE& Indonesia \\
& 13& KLSE&Malaysia  \\
& 14& N225&Japan  \\
& 15& STI& Singapore \\
& 16& KS11&Korea  \\
& 17& TWII&Taiwan  \\
\hline
Europe& 18&ATX & Austria \\
& 19&BFX & Belgium \\
& 20&FCE.NX &France  \\
& 21& GDAXI& Germany \\
& 22&AEX & Holland \\
& 23&MIBTEL &Italy  \\
& 24&SSMI &Switzerland  \\
& 25&FTSE &U.K.
\\\hline
\end{tabular}%
\caption{List of 25 markets. We obtain data from the website
http://finance.yahoo.com.} \label{table:market}
\end{table}

Through transfer entropy, this paper focuses quantitatively on the
direction of information flow between 25 stock markets to
determine which market serves as a source of information for
global stock indices. Drawn from the economic system, a plethora
of empirical data reflecting economic conditions can be obtained.
The time series of a composite stock price index provides prime
data accurately reflecting economic conditions. Therefore, we
analyzed the daily time series of the 25 stock indices listed in
Table \ref{table:market} for the period of 2000 -- 2007 using
transfer entropy in order to examine the information flow between
stock markets and identify the hub.

\section{Transfer entropy}

Transfer entropy, which measures the directionality of a variable
with respect to time was recently introduced by Schreiber
\cite{schreiber} based on the probability density function (PDF).
Let us consider two discrete and stationary process, $I$ and $J$.
Transfer entropy relates $k$ previous samples of process $I$ and
$l$ previous samples of process $J$ and is defined as follows:
\begin{equation}
T_{J \rightarrow I} = \sum p(i_{t+1}, i_t^{(k)}, j_t^{(l)}) \log
\frac{p( i_{t+1} \mid i_t^{(k)}, j_t^{(l)})}{p( i_{t+1} \mid
i_t^{(k)})},
\end{equation}
where $i_t$ and $j_t$ represent the discrete states at time $t$ of
$I$ and $J$, respectively. $i_t^{(k)}$ and $j_t^{(l)}$ denote $k$
and $l$ dimensional delay vectors of two time consequences $I$ and
$J$, respectively. The joint PDF $p(i_{t+1}, i_t^{(k)},
j_t^{(l)})$ is the probability that the combination of $i_{t+1}$,
$i_t^{(k)}$ and $j_t^{(l)}$ have particular values. The
conditional PDF $p(i_{t+1} \mid  i_t^{(k)}, j_t^{(l)})$ and $p(
i_{t+1} \mid i_t^{(k)})$ are the probability that $i_{t+1}$ has a
particular value when the value of previous samples $i_t^{(k)}$
and $j_t^{(l)}$ are known and $i_t^{(k)}$ are known, respectively.

The transfer entropy with index $J \rightarrow I$ measures the
extent to which the dynamics of process $J$ influences the
transition probabilities of another process $I$. Reverse
dependency is calculated by exchanging $i$ and $j$ of the joint
and conditional PDFs. The transfer entropy is explicitly
asymmetric under the exchange of $i_t$ and $j_t$. It can thus
provide information regarding the direction of interaction between
two time series.

The transfer entropy is quantified by information flow from $J$ to
$I$. The transfer entropy can be calculated by subtracting the
information obtained from the last observation of $I$ exclusively
from the information about the latest observation $I$ obtained
from the final joint observation of $I$ and $J$. This is the nexus
of transfer entropy. Therefore, transfer entropy can be rephrased
as
\begin{equation}
T_{J \rightarrow I} = h_I(k) - h_{IJ}(k,l),\label{eq:TE}
\end{equation}
where
\begin{eqnarray}
h_I(k)=-\sum p(i_{t+1}, i_t^{(k)}) \log p( i_{t+1} \mid i_t^{(k)}) \\
h_{IJ}(k,l)=-\sum p(i_{t+1}, i_t^{(k)}, j_t^{(l)}) \log p( i_{t+1}
\mid i_t^{(k)}, j_t^{(l)}).
\end{eqnarray}

\section{Empirical data analysis} \label{data}

We analyzed the daily data records of 25 stock exchange indices
for the period January 2000 -- December 2007. We adopted the log
returns to describe a financial time series $x(n)\equiv \ln (S(n))
- \ln (S(n-1))$. Where $S_n$ means the index of $n$-th trading
day. We partitioned the real value $x(n)$ into discretized value
$A(n)$; $A(n) = 0$ for $x(n) \leq -d$ (decrease), $A(n) = 1$ for $
-d < x(n) < d$ (intermediate), $A(n) = 2$ for $x(n) \geq d/2$
(increase). We set $d=0.04$ in our analysis. At that value, the
probabilities of three states are approximately same for all
indices. Additionally, we set $k=l=1$.

\begin{figure}[tbph]
\begin{center}
\mbox{
        \subfigure[]
              {\scalebox{0.16}
       {\hspace{0mm}\epsfig{file=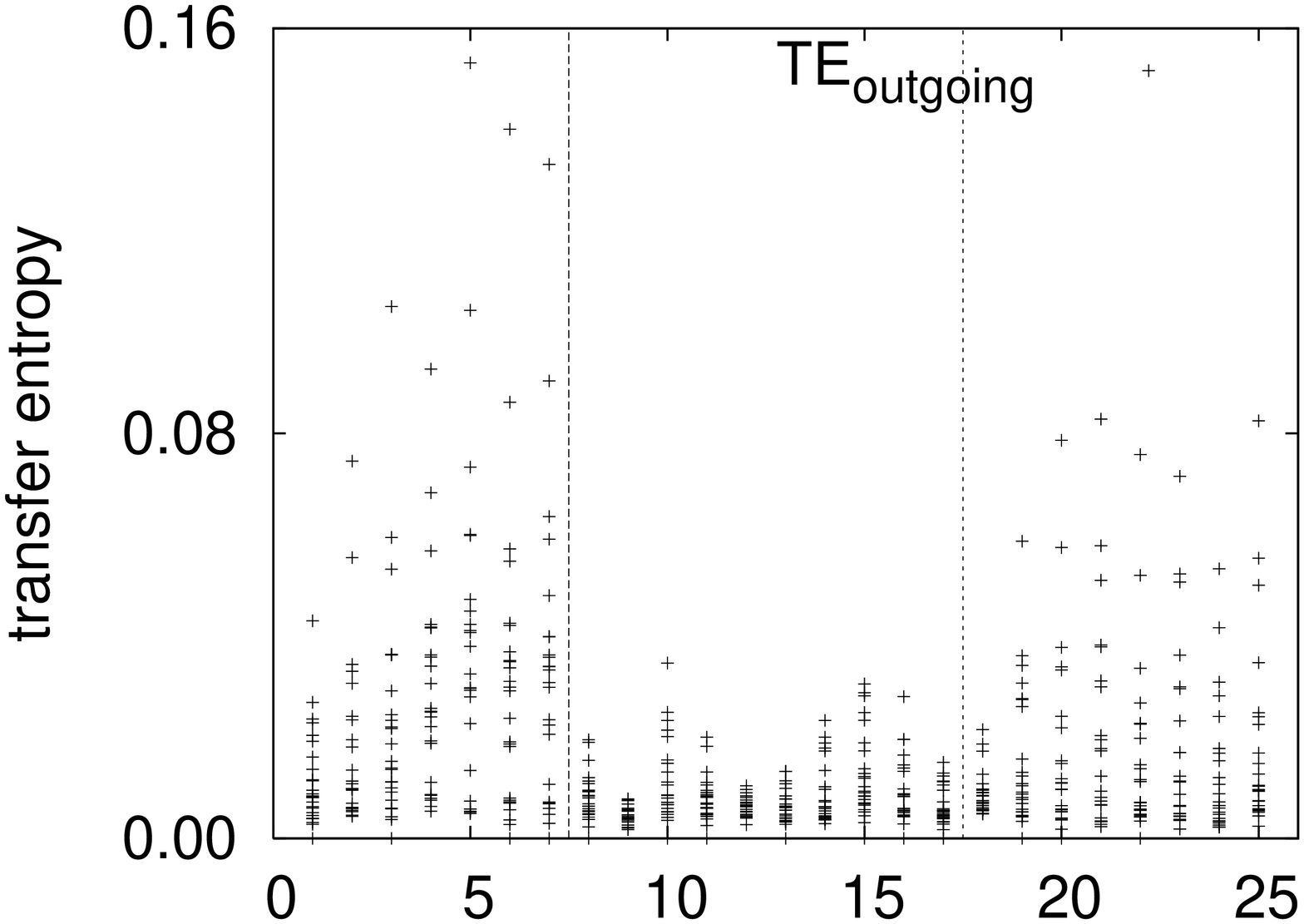,angle=-90}}
      }}
\mbox{
        \subfigure[]
              {\scalebox{0.16}
       {\hspace{0mm}\epsfig{file=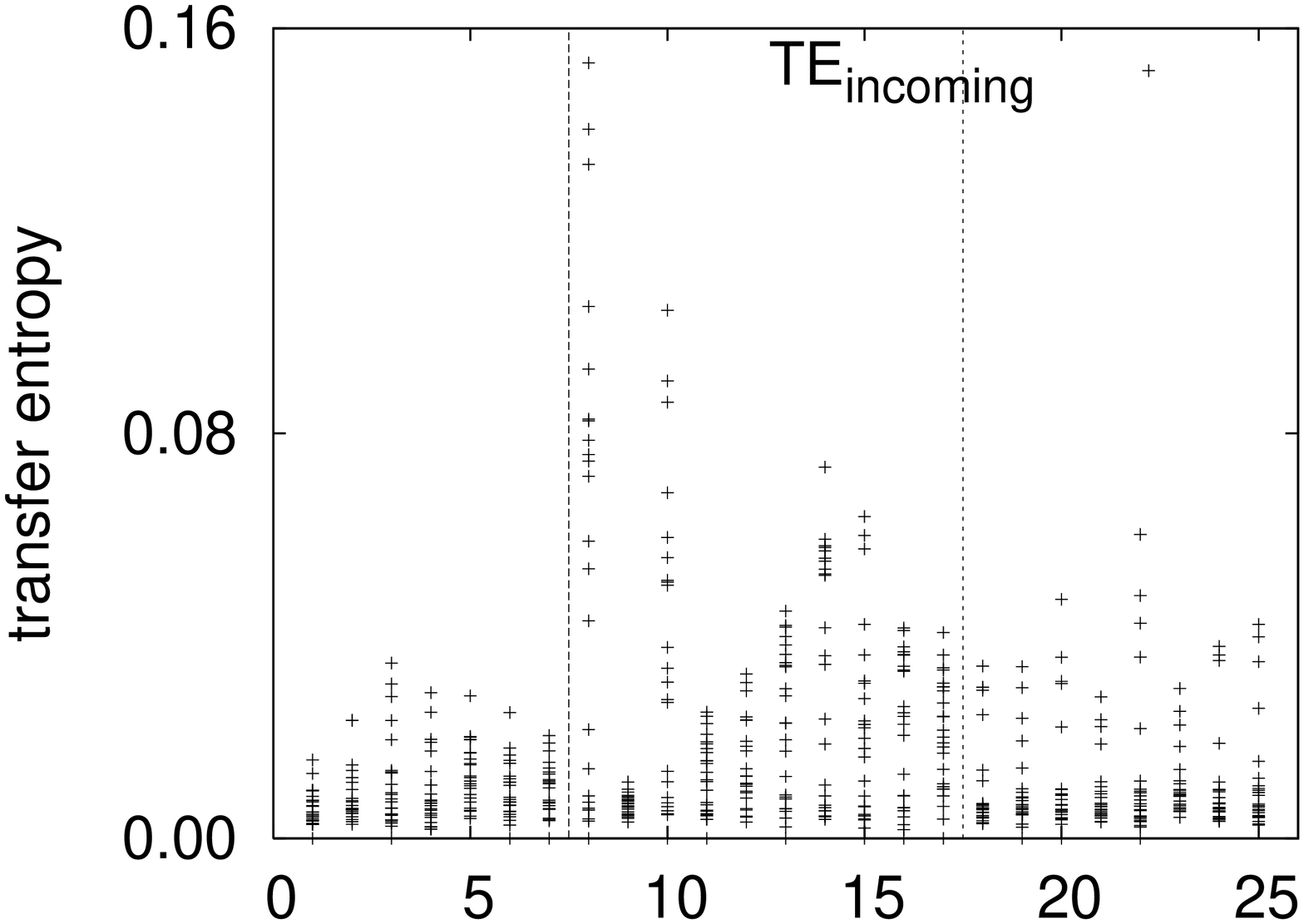,angle=-90}}
      }}
\mbox{
        \subfigure[]
              {\scalebox{0.16}
       {\hspace{0mm}\epsfig{file=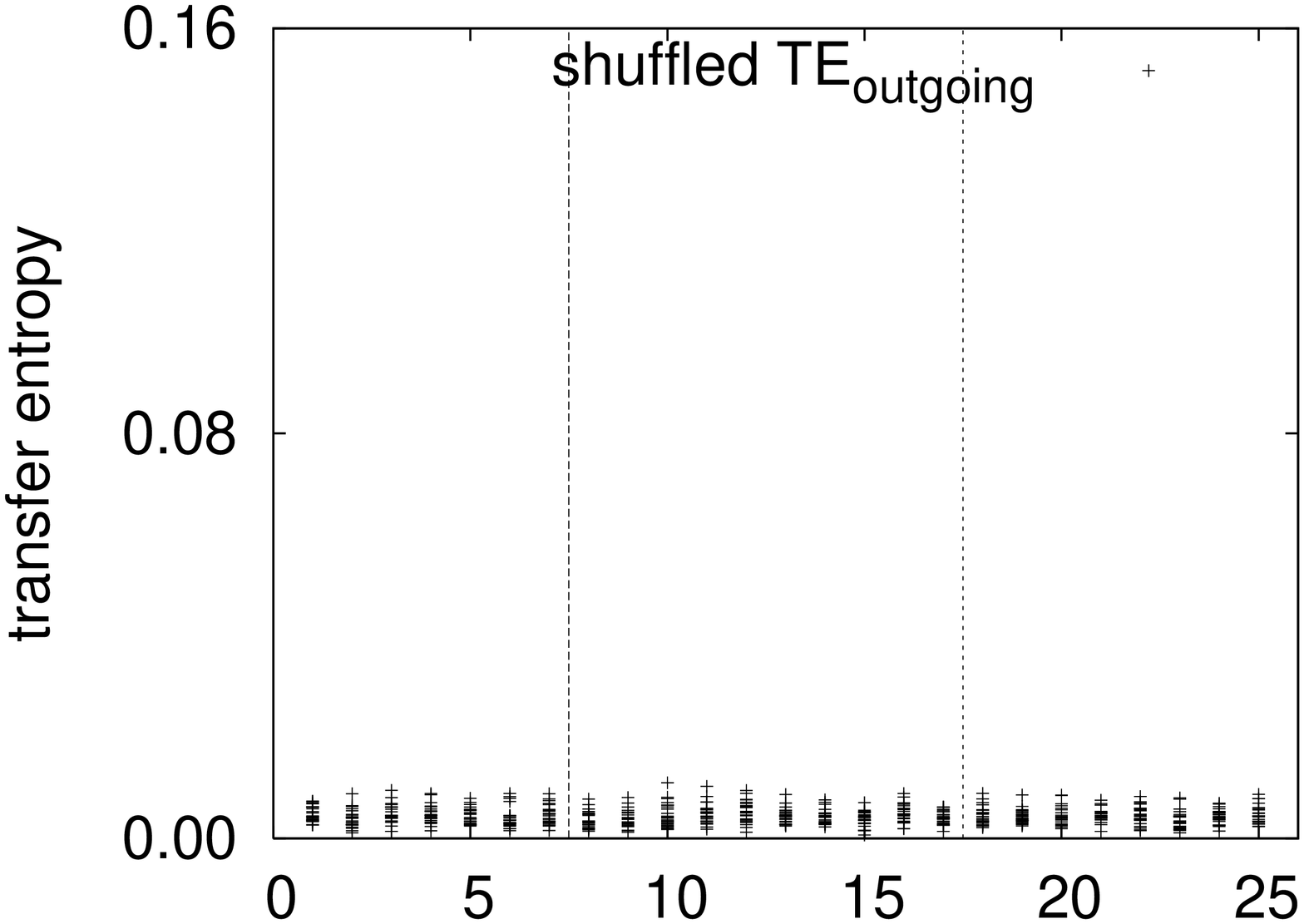,angle=-90}}
      }}
      \mbox{
        \subfigure[]
              {\scalebox{0.16}
       {\hspace{0mm}\epsfig{file=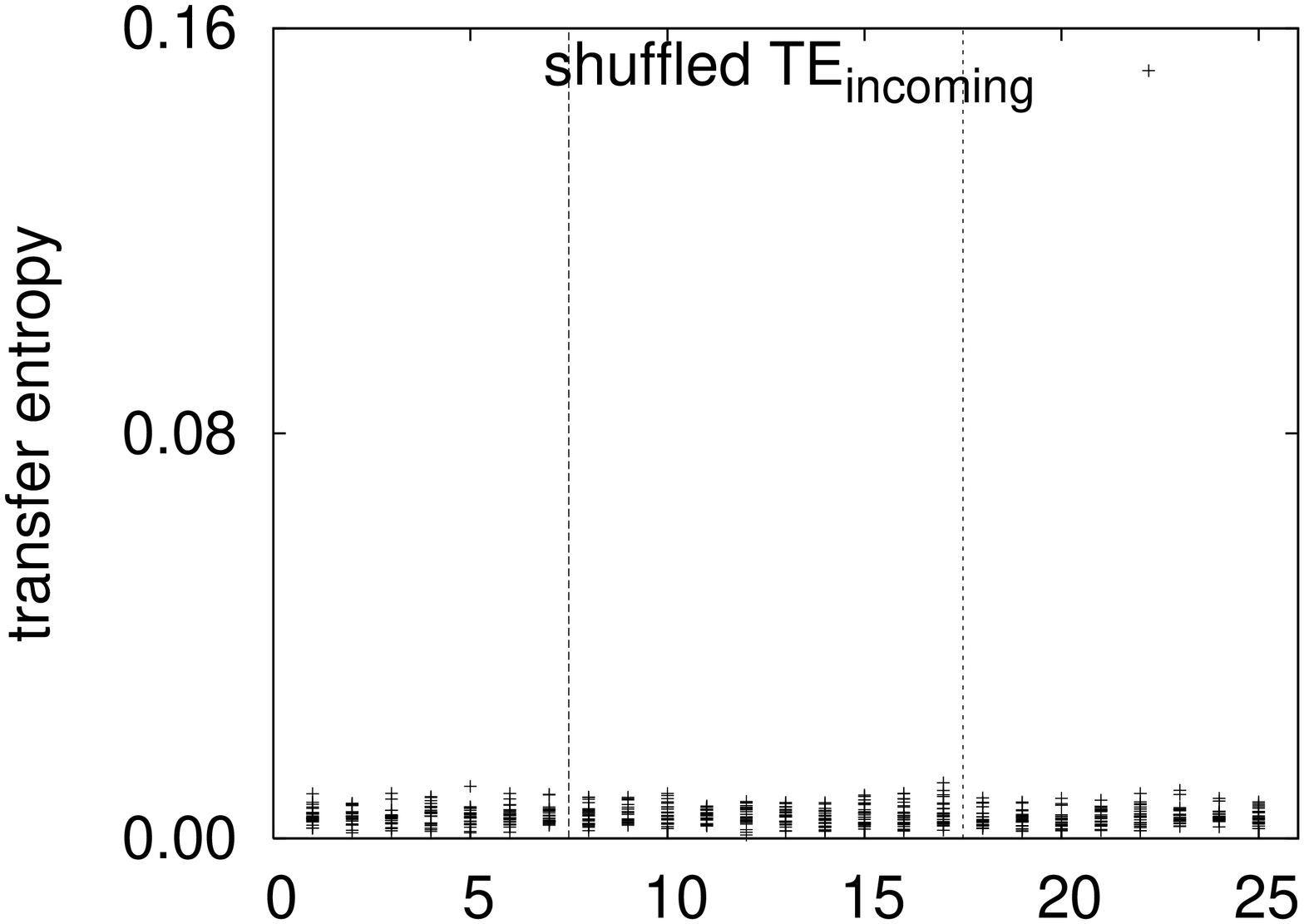,angle=-90}}
      }}
\end{center}
\par
\vspace{-5mm} \caption{Plots of (a) the outgoing and (b) the
incoming transfer entropy. Plots of (c) the outgoing and (d) the
incoming transfer entropy for the shuffled data. Numbers in
$x$-axis indicate markets listed in Table
\ref{table:market}.}\label{fig:te}
\end{figure}

Fig. \ref{fig:te} is the plots of the outgoing and the incoming
transfer entropy. Numbers in $x$-axis indicate markets listed in
Table \ref{table:market} while the dotted lines in the graphs
serve to distinguish the continents. Each stock market interacts
with 24 other stock markets, and the transfer entropy for all
possible pairs can be calculated. Fig. \ref{fig:te}(a) represents
the transfer entropy from the market indicating $x$-axis to the
other 24 markets, that is, the outgoing transfer entropy. Fig.
\ref{fig:te}(b) represents the transfer entropy from the other 24
markets to the market in $x$-axis, that is, the incoming transfer
entropy. The markets in the Americas have high values of outgoing
transfer entropy, while the value of the transfer entropy for the
markets in Asia remains low. Moreover, the incoming transfer
entropy for Asia is higher than that of the Americas and Europe.
From Fig. \ref{fig:te}, we can determine that the directionality
of influence for a market index is generally from the Americas to
Asia. To clarify the tendencies of information flow, we shuffled
the data and verified the transfer entropy. In Fig.
\ref{fig:te}(c) and Fig. \ref{fig:te}(d), the transfer entropy for
the shuffled data is generally approaching zero, information flow
between markets has nearly disappeared.

\begin{figure}[tbph]
\begin{center}

              {\scalebox{0.75}
       {\hspace{-5mm}\epsfig{file=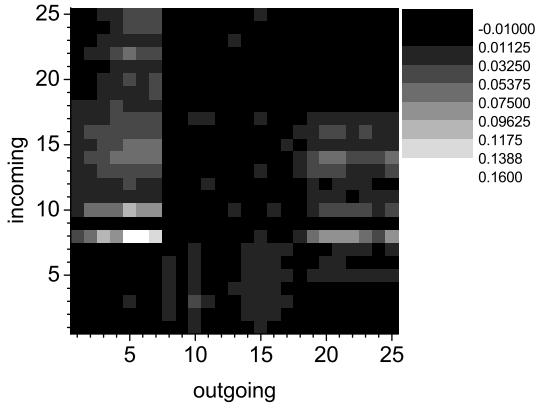,angle=0}}
      }
             \end{center}
\par
\vspace{0mm} \caption{Gray-scale map of the transfer entropy
between stock markets.} \label{fig:matrix}
\end{figure}

Fig. \ref{fig:matrix} is the gray-scale map of the transfer
entropy between stock markets using the same data as in Fig.
\ref{fig:te}. The direction of information flow is from the
$x$-axis to $y$-axis. The darker the lattice, the lower the
transfer entropy. The markets in the Americas influence the
markets in both Asia and Europe, but influence on the Asian
markets is particularly strong. The European markets also
influence the Asian markets, but to a lesser degree than do the
American markets.

\begin{figure}[tbph]
\begin{center}
              {\scalebox{0.75}
       {\hspace{-5mm}\epsfig{file=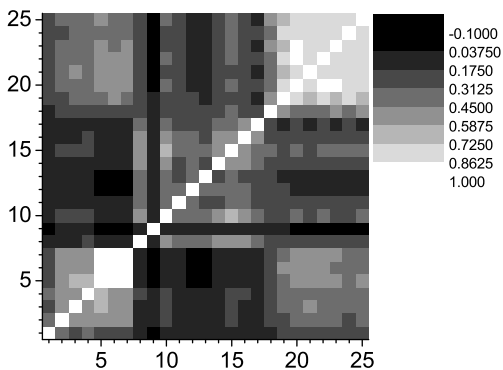,angle=0}}
      }
\end{center}
\par
\vspace{0mm} \caption{Gray-scale map of the cross-correlations
between stock markets.} \label{fig:corrmat}
\end{figure}

Fig. \ref{fig:corrmat} is the gray-scale map of the
cross-correlations between stock markets from log-return time
series $x(n)$. The cross-correlation does not represent
directionality between markets. Therefore, the graph is
symmetrical, centering around the diagonal. In this graph, a light
lattice means that the cross-correlation is high. The American and
European markets are highly correlated between themselves, while
the Asian markets are not when compared with the American and
European markets. Also, the American and Asian markets are rarely
correlated with each other. From these results, we can induce (or
that the comparatively mature American and European markets are
well clustered between themselves, while the relatively emerging
Asian markets are not. Furthermore, between the American and Asian
markets there exists low cross-correlation and high transfer
entropy, demonstrating an imbalance of information flow.

\begin{figure}[tbph]
\begin{center}
\mbox{
        \subfigure[]
              {\scalebox{0.4}
       {\hspace{0mm}\epsfig{file=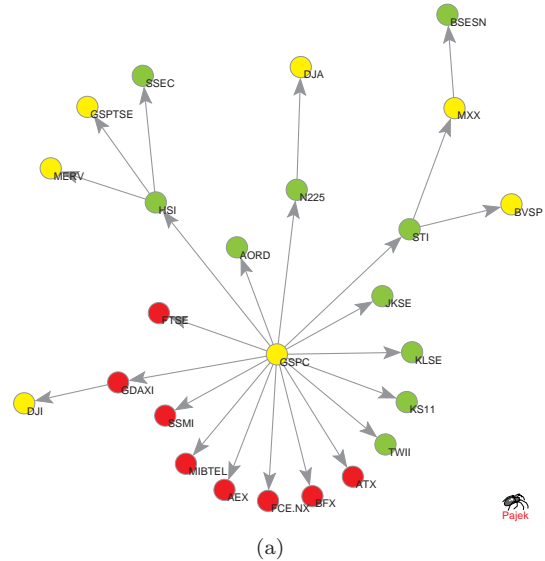,angle=0}}
      }}
\mbox{
        \subfigure[]
              {\scalebox{0.4}
       {\hspace{0mm}\epsfig{file=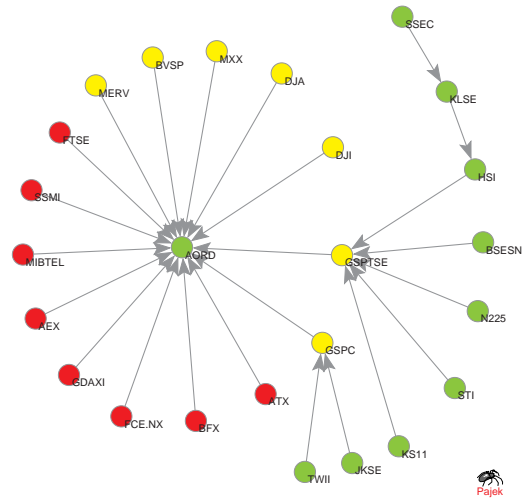,angle=0}}
      }
    }
\end{center}
\par
\vspace{-5mm} \caption{Minimum spanning tree for (a) the outgoing
transfer entropy and (b) the incoming transfer entropy. The
minimum spanning tree is drawn by Pajek
(http://vlado.fmf.uni-lj.si/pub/networks/pajek/).} \label{fig:mst}
\end{figure}

To observe schematically the relation of information flow between
the stock markets, the minimum spanning tree is depicted in Fig.
\ref{fig:mst}. The colors of nodes indicate the continent housing
the markets. The minimum spanning tree is constructed by
connecting the highest outgoing (Fig. \ref{fig:mst}(a)) and
incoming (Fig. \ref{fig:mst}(b)) transfer entropy. In Fig.
\ref{fig:mst}(a), the GSPC is located at the focus of the network,
connected with the Asian and European markets. The GSPC plays the
role of information source among stock markets. On the other hand,
in Fig. \ref{fig:mst}(b) the Australian market, AORD, is located
in the center of the minimum spanning tree for incoming transfer
entropy.

\section{Conclusions} \label{conclusion}

We observed the transfer entropy and the cross-correlation of the
daily data of 25 stock markets to investigate the intensity and
the direction of information flow between stock markets. The value
of the outgoing transfer entropy from the American markets is
high, as it is from the European markets. The information flow
mainly streams to the Asia/Pacific region. However, the transfer
entropy between markets on the same continent is not high. We also
reveal that markets on the same continent have higher values of
cross-correlation for log-return. This indicates that
intra-continental market indices fluctuate simultaneously with
common deriving mechanisms. The cross-correlation between America
and Europe is relatively high. Therefore, we can see that American
and European stock markets fluctuate in tune with common deriving
mechanisms. However, the Asia/Pacific markets fluctuate
simultaneously with other common deriving mechanisms.
Consequently, we can predict that the deriving mechanism for
America and Europe does not depend on the Asia/Pacific market
situation but rather that Asia is affected by the American and
European market situation. From Fig. \ref{fig:mst}, we can confirm
the above description. The GSPC is located in the center of the
star network, influencing all the other markets. Therefore, the
Asian markets represent similar patterns of time series, although
they do not actively interact  with each other.

Transfer entropy is a proper measure for the investigation of the
information flow between global stock markets. Through this
measure we can quantify information transportation between
underlying mechanisms that govern stock markets even though we
cannot concretely identify the mechanism.

\acknowledgments

This work is supported in part by the Creative Research
Initiatives of the Korea Ministry of Science and Technology and
the Second Brain Korea 21 project.


\end{document}